\documentclass[runningheads]{llncs}

\usepackage[T1]{fontenc}
\usepackage{graphicx}
\usepackage{syntax} 
\usepackage{amsmath}
\usepackage{amssymb}
\usepackage{tcolorbox}
\usepackage{listings}
\usepackage[noend]{algpseudocode}
\usepackage{algorithm}
\usepackage{pgfplots}
\usepackage{subcaption}
\usepackage{mathpartir}
\usepackage{xcolor}

\usepackage{tikz}
\usetikzlibrary{calc}
\usetikzlibrary{arrows.meta}

\usepackage{hyperref}
\usepackage{color}

\definecolor{codegreen}{rgb}{0,0.6,0}
\definecolor{codegray}{rgb}{0.5,0.5,0.5}
\definecolor{codepurple}{rgb}{0.58,0,0.82}
\definecolor{backcolour}{rgb}{0.95,0.95,0.92}

\lstdefinestyle{vehicleStyle}{
    backgroundcolor=\color{backcolour},   
    commentstyle=\color{codegreen},
    keywordstyle=\color{magenta},
    numberstyle=\tiny\color{codegray},
    stringstyle=\color{codepurple},
    basicstyle=\ttfamily\footnotesize,
    breakatwhitespace=false,         
    breaklines=true,                 
    captionpos=b,                    
    keepspaces=true,                 
    numbers=left,                    
    numbersep=5pt,                  
    showspaces=false,                
    showstringspaces=false,
    showtabs=false,                  
    tabsize=2
}

\algrenewcommand\algorithmicindent{1.0em}

\algnewcommand{\algorithmictraverse}{\textbf{traverse}}
\algdef{SE}[TRAVERSE]{Traverse}{EndTraverse}[1]
{
    \algorithmictraverse\ #1\ \algorithmicof
}
{
    \algorithmicend\ \algorithmictraverse
}
\algtext*{EndTraverse}%

\algnewcommand{\algorithmicswitch}{\textbf{switch}}
\algdef{SE}[SWITCH]{Switch}{EndSwitch}[1]
{
    \algorithmicswitch\ #1\ \algorithmicof
}
{
    \algorithmicend\ \algorithmicswitch
}
\algtext*{EndSwitch}%

\algnewcommand{\algorithmiccase}{\textbf{case}}
\algnewcommand{\algorithmicof}{\textbf{of}}
\algdef{SE}[CASE]{Case}{EndCase}[1]{
    \algorithmiccase\ #1 $\rightarrow$
}
{
    \algorithmicend\ \algorithmiccase
}%
\algtext*{EndCase}%

\algdef{SE}[DEFAULT]{Default}{EndDefault}{\algorithmicdefault}{\algorithmicend\ \algorithmicdefault}%

\algnewcommand{\algorithmicthrow}{\textbf{throw}}
\algnewcommand{\Throw}[1]{\algorithmicthrow~#1}

\algnewcommand{\algorithmicnote}{\textbf{note}}
\algnewcommand{\Note}[1]{\algorithmicnote~#1}

\algblock[TryCatchFinally]{try}{EndTry}
\algcblock[TryCatchFinally]{TryCatchFinally}{Finally}{EndTry}
\algcblockdefx[TryCatchFinally]{TryCatchFinally}{Catch}{EndTry}
	[1]{\textbf{catch} #1}
	{}
\algtext*{EndTry}

\newcommand{\compileSymbol}{\textsc{compile}}
\newcommand{\compile}[1]{\compileSymbol(#1)}

\newcommand{\compileBodySymbol}{\textsc{compileBody}}
\newcommand{\compileBody}[1]{\compileBodySymbol(#1)}

\newcommand{\compileBodyRecSymbol}{\textsc{compileBodyAndEqualities}}
\newcommand{\compileBodyRec}[1]{\compileBodyRecSymbol(#1)}

\newcommand{\compileComparisonSymbol}{\textsc{compileComparison}}
\newcommand{\compileComparison}[3]{\compileComparisonSymbol(#1,\,#2,\,#3)}

\newcommand{\elimExistsSymbol}{\textsc{eliminateExists}}
\newcommand{\elimExists}[2]{\elimExistsSymbol(#1,\,#2)}

\newcommand{\elimNotSymbol}{\textsc{lowerNot}}
\newcommand{\elimNot}[1]{\elimNotSymbol(#1)}

\newcommand{\purifySymbol}{\textsc{compileArith}}
\newcommand{\purify}[2]{\purifySymbol(#1,\,#2)}

\newcommand{\elimVarSymbol}{\textsc{eliminateUserVariable}}
\newcommand{\elimVar}[2]
{\elimVarSymbol(#1,\,#2)}

\newcommand{\evalAtSymbol}{\textsc{evalAtIndex}}
\newcommand{\evalAt}[2]
{\evalAtSymbol(#1,\,#2)}

\newcommand{\evalTrivialComparisonSymbol}{\textsc{evalComp}}
\newcommand{\evalTrivialComparison}[3]
{\evalTrivialComparisonSymbol(#1,\,#2,\,#3)}

\newcommand{\optimiseQueriesSymbol}{\textsc{compileProtoqueries}}
\newcommand{\optimiseQuery}[1]{\optimiseQueriesSymbol(#1)}

\newcommand{\eliminateEqualAppsSymbol}{\textsc{optimiseNetworkApplications}}
\newcommand{\eliminateEqualApps}[1]{\eliminateEqualAppsSymbol(#1)}

\newcommand{\findConnectedComponentsSymbol}{\textsc{equivClasses}}
\newcommand{\findConnectedComponents}[4]{\findConnectedComponentsSymbol(#1,\,#2,\,#3,\,#4)}

\newcommand{\equalitySolverSymbol}{\textsc{equalitySolver}}
\newcommand{\equalitySolver}[2]
{\equalitySolverSymbol(#1, #2)}

\newcommand{\reduceDimsSymbol}{\textsc{ReduceToZeroDimAssertions}}
\newcommand{\reduceDims}[1]{\reduceDimsSymbol{}(#1)}

\newcommand{\andProtoqueriesSymbol}{\textsc{Conjunct}}
\newcommand{\andProtoqueries}[2]{\andProtoqueriesSymbol{}(#1, #2)}

\newcommand{\orProtoqueriesSymbol}{\textsc{Disjunct}}
\newcommand{\orProtoqueries}[2]{\orProtoqueriesSymbol{}(#1, #2)}

\newcommand{\bool}{\mathbb{B}}
\newcommand{\nat}{\mathbb{N}}
\newcommand{\real}{\mathbb{R}}
\newcommand{\rationals}{\mathbb{Q}}
\newcommand{\numType}{\mathbb{D}}

\newcommand{\emptyList}{\varnothing}
\newcommand{\cons}[2]{#1 :: #2}

\newcommand{\bop}[3]{#1\,#2\,#3}

\newcommand{\boolSymbol}{\texttt{Bool}}
\newcommand{\realSymbol}{\texttt{Real}}

\newcommand{\tensorSymbol}{\texttt{Tensor}}

\newcommand{\andSymbol}{\texttt{and}}
\newcommand{\orSymbol}{\texttt{or}}

\newcommand{\notSymbol}{\texttt{not}}
\newcommand{\addSymbol}{\texttt{+}}
\newcommand{\mulSymbol}{\texttt{*}}
\newcommand{\eqSymbol}{\texttt{==}}
\newcommand{\neqSymbol}{\texttt{!=}}
\newcommand{\leqSymbol}{\texttt{<=}}
\newcommand{\leSymbol}{\texttt{<}}
\newcommand{\geSymbol}{\texttt{>}}
\newcommand{\geqSymbol}{\texttt{>=}}
\newcommand{\forallSymbol}{\texttt{forall}}
\newcommand{\existsSymbol}{\texttt{exists}}

\newcommand{\gtype}{\ensuremath{\tau}}
\newcommand{\gexpr}{\ensuremath{e}}
\newcommand{\gvar}{\ensuremath{x}}

\newcommand{\hBinder}[2]{(\hVar{#1} : #2)}

\newcommand{\hVar}[1]{\ensuremath{\texttt{#1}}}

\newcommand{\hRealType}{\ensuremath{\realSymbol{}}}

\newcommand{\hAdd}[2]{\bop{#1}{\addSymbol{}}{#2}}
\newcommand{\hMul}[2]{\bop{#1}{\mulSymbol{}}{#2}}

\newcommand{\hBoolType}{\ensuremath{\boolSymbol{}}}

\newcommand{\hNot}[1]{\ensuremath{\notSymbol{} \: #1}}
\newcommand{\hAnd}[2]{\bop{#1}{\andSymbol{}}{#2}}
\newcommand{\hOr}[2]{\bop{#1}{\orSymbol{}}{#2}}

\newcommand{\hEq}[2]{\bop{#1}{\eqSymbol{}}{#2}}

\newcommand{\hLeq}[2]{\bop{#1}{\leqSymbol{}}{#2}}

\newcommand{\hQuantify}[4]{
\ensuremath{#1 \: \hBinder{#2}{#3} \texttt{.} \: #4}}
\newcommand{\hForall}[3]{\hQuantify{\forallSymbol{}}{#1}{#2}{#3}}
\newcommand{\hExists}[3]{\hQuantify{\existsSymbol{}}{#1}{#2}{#3}}

\newcommand{\hTensorType}[2]{\ensuremath{\ensuremath{\tensorSymbol{}} \: #2}}

\newcommand{\hAt}[2]{#1[#2]}
\newcommand{\hSeq}[1]{\ensuremath{[#1]}}

\newcommand{\hRealTensorType}[1]{\hTensorType{\hRealType}{#1}}
\newcommand{\hNetCtx}{N}

\newtcolorbox{authorComment}[1]{colback=#1}
\begin{document}

\title{Compiling High-Level Neural Network Specifications into VNN-LIB Queries}
\titlerunning{Efficient Compilation to Neural Network Solvers}

\author{Matthew L. Daggitt \inst{1,2}\orcidID{0000-0002-2552-3671} \and
Wen Kokke\inst{2,3}\orcidID{0000-0002-1662-0381} \and
Robert Atkey\inst{4}\orcidID{0000-0002-4414-5047}}
\authorrunning{M.L. Daggitt et al.}

%\author{}

\institute{
University of Western Australia, Perth, Australia \\\email{matthewdaggitt@gmail.com} 
\and
Heriot-Watt University, Edinburgh, Scotland
\and
Well-Typed, London, UK
\and
University of Strathclyde, Glasgow, Scotland
}
%\institute{}

\maketitle

\vspace{-2em}

\begin{abstract}
The formal verification of traditional software has been revolutionised by verification-orientated languages such as Dafny and F* which enable developers to write high-level specifications that are automatically compiled down to low-level SMT-LIB queries. In contrast, neural network verification currently lacks such infrastructure, often requiring users to express requirements in formats close to the low-level VNN-LIB query format. This gap persists because targeting VNN-LIB presents unique algorithmic challenges when compared to targeting SMT-LIB: VNN-LIB is restricted to a fixed finite set of variables representing the input and outputs of the network, and even toy neural network specifications have an extremely large number of variables.
    
In this paper, we present the first algorithm for compiling high-level neural network specifications into optimised VNN-LIB queries. Our algorithm is numerically sound and supports a far rich logical fragment than existing tools, including transformations of variables, first-class quantifiers, and specifications involving multiple networks or multiple applications of the same network. 
We implement this algorithm within the Vehicle framework and demonstrate that its performance is asymptotically optimal for benchmark specifications.

%\keywords{Verification \and Neural Network Solvers \and DSL compilation.}
\end{abstract}

\section{Introduction}

In conventional software engineering the gap between high-level requirements and automated verification has been successfully bridged by domain specific languages (e.g. Rosette~\cite{torlak2013growing} and Cryptol~\cite{lewis2007cryptol}) and general-purpose verification-oriented programming languages (e.g. Dafny~\cite{leino2010dafny}, Liquid Haskell~\cite{vazou2016liquid} and F*~\cite{swamy2016dependent}). 
These languages allow developers to express specifications using high-level logical constructs such as quantifiers, implications, and function abstraction. 
A specialised compiler then reduces these rich specifications into low-level SMT-LIB queries~\cite{barrett2010smt}, which are automatically discharged by satisfiability modulo theories (SMT) solvers (e.g. Z3~\cite{moura2008z3}). This infrastructure has been pivotal in making formal methods accessible to non-experts.

In theory, a similar methodology should apply to the verification of neural networks. Users should be able to write high-level logical specifications for network behaviour which are then compiled down to VNN-LIB~\cite{vnnlib2025} -- the standardised low-level query format for neural network verifiers -- and then discharged by neural network solvers~\cite{ivanov2019verisig,lopez2023nnv,muller2022prima,NCubeV,wang2021beta,wu2024marabou}.
However, there are two key differences between VNN-LIB and SMT-LIB. 
First, VNN-LIB queries are not self-contained: they explicitly reference the neural network model as an external artefact. 
Second, to scale to non-trivial networks, most neural network solvers rely on abstract-interpretation–based techniques. Due to the latter, VNN-LIB deliberately restricts queries to a fixed, finite set of variables corresponding to the network’s direct inputs and outputs.
Consequently, constructing a general-purpose compiler that targets VNN-LIB presents unique challenges that do not exist in the standard SMT-based workflow.

\subsection{Motivating example}
\label{sec:motivating-example}

The following specification over a network ${f : \real^2 \rightarrow \real}$ illustrates the problem:
\begin{lstlisting}{language=Python}
exists (a : Tensor Real [2]). 
  a[0] > 0 and a[1] > 0 and f([ a[0]+a[1], a[0]-a[1] ]) > 0
\end{lstlisting}
When targeting SMT-LIB, a compiler can declare an arbitrary number of SMT-LIB variables and this allows a direct translation of the above to:
\begin{lstlisting}
(declare-const a0 Real) ; first element in tensor a
(declare-const a1 Real) ; second element in tensor a
(declare-const x0 Real) ; first input to f
(declare-const x1 Real) ; second input to f
(declare-const y  Real) ; output of f

... ; Encoding of f in terms of x0, x1 and y

(assert (= x0 (+ a0 a1)))
(assert (= x1 (- a0 a1)))
(assert (and (> a0 0) (> a1 0) (> y 0)))
\end{lstlisting}
In this paper we will refer to the quantified variables in the original specification (e.g. \texttt{a}, \texttt{a0}, \texttt{a1}) as \emph{user variables} and the variables that represent the inputs and outputs of the neural network (e.g. \texttt{x0}, \texttt{x1}, \texttt{y}) as \emph{network variables}.

In contrast, VNN-LIB restricts queries to only use network variables and therefore a compiler cannot introduce variables representing the user variables. Instead, it must solve the simultaneous equations $x_0 = a_0 + a_1$ and $x_1 = a_0 - a_1$ to obtain the solutions $a_0 = 0.5 (x_0 + x_1)$ and $a_1 = 0.5 (x_0 - x_1)$ for the user variables and then generate VNN-LIB 2.0 code of the form:
\begin{lstlisting}
(declare-network f
  (declare-input  x Real [2])
  (declare-output y Real []))

(assert (and 
  (> (* 0.5 (+ x0 x1)) 0) (> (* 0.5 (- x0 x1) 0)) (> y 0)))
\end{lstlisting}
Consequently, compiling a high-level specification to VNN-LIB is not merely a syntactic translation; in the general case, it requires a non-trivial transformation from the user variables in the high-level specification down to the limited set of network variables available in VNN-LIB.
This transformation between the two sets of variables has been referred to as the \emph{embedding gap} in~\cite{cordeiro2025neural}.

The scale of neural network verification problems further increases the complexity of the transformation. As the example above shows, in the general case, the user variables can only be solved in terms of the network variables when working at the level of the individual elements of the tensors. 
However, even simple Gaussian elimination is $O(n^3)$ in the worst-case, and in the toy MNIST problem each input tensor consists of hundreds of inputs ($n = 784$). This makes naive variable elimination algorithms impractical for even trivial verification problems. 
The algorithmic complexity is increased yet again when one considers specifications that contain multiple networks (e.g. global robustness~\cite{kabaha2024verification}) and chained network applications (e.g. encoder-decoder models~\cite{kingma2019introduction}).

\subsection{Existing specification languages}
\label{sec:existing-languages}

Due to this complexity, there has been comparatively little progress in building high-level specification languages for neural network verification. Notable examples of domain-specific languages include the Python-based DNNV framework~\cite{shriver2021dnnv} and the WhyML-based CAISAR framework~\cite{girardsatabin2022caisar}. However their compilation algorithms are unpublished and, crucially, they avoid the complexity described in Section~\ref{sec:motivating-example} by either relying on templates, where users specify parameters to pre-defined specifications, or restricting their specification language to subsets of logic that map directly to the underlying verifier's capabilities. As a result, they suffer from the following expressivity limitations:
\begin{enumerate}
    \item \textbf{Network inputs}: User variables must appear directly as network inputs, or be subject only to a restricted class of transformations. As a result, specifications cannot in general include data normalisation steps, preventing users from writing properties over semantically meaningful, unnormalised inputs.
        
    \item \textbf{Restricted syntax}: Quantifiers are not first-class constructs in the language.
    This hinders the ability of users to naturally express more complex specifications (e.g. quantifying over multiple subrows of an input tensor).
    
    \item \textbf{Single network applications}: Specifications can only contain a single application of a single network, and therefore cannot express hyper-properties involving multiple applications of a network (e.g., monotonicity~\cite{liu2020certified}) or relationships between different networks (e.g., teacher-student distillation~\cite{park2021learning})\footnote{The community works around this limitation by merging multiple networks into a single ONNX file for verification. However, this approach has significant drawbacks: the user is no longer verifying the same artifact they are deploying, and the solver loses the ability to share information between multiple copies of the same network.}.
\end{enumerate}
In short, current high-level specification languages fail to provide a meaningful abstraction over neural network solvers. By exposing solver limitations directly in the language design, they unnecessarily constrain users to a narrow fragment of the specifications they may wish to write.

\subsection{Contributions}

This paper addresses this gap by providing the first description of an  algorithm capable of compiling high-level specifications written in a fragment of first-order logic (FOL) extended with tensors to equisatisfiable trees of VNN-LIB queries.
This fragment of FOL supports the following novel features not found in any other neural network specification language:
\begin{enumerate}
    \item Quantified user variables can be near-freely transformed before being used as inputs to neural networks.
    \item Quantifiers are first-class constructs in the language.
    \item The specification can contain arbitrarily nested applications of an arbitrary number of neural networks.
    \item The specification can contain non-linear constraints on user variables\footnote{Neural network solvers that support non-linear constraints are currently rare~\cite{NCubeV}.}.
\end{enumerate}
Furthermore, under the assumption that the underlying algebraic solvers used by the algorithm are sound, the algorithm is sound over any numeric type, e.g. real or floating point.
We implement this algorithm within the existing Vehicle framework~\cite{daggitt2025vehicle} and demonstrate that compilation time scales linearly in the size of the input tensors of the neural networks for a set of representative specifications.

Our algorithm therefore succeeds in abstracting over the limitations of existing neural network solvers and is an important step in building more accessible interfaces to neural network solvers that are suitable for wide-spread adoption.

\subsection{Limitations}
\label{sec:limitations}

The fragment of FOL supported by our algorithm only imposes two expressivity restrictions compared to full FOL, both of which are unavoidable consequences arising from current solver technology:
\begin{enumerate}
    \item \textbf{Non-alternating quantifiers:} The specification cannot contain nested alternating existential and universal quantifiers. This restriction arises as both solvers and the syntax of VNN-LIB only support satisfiability queries. 
    \item \textbf{Solver-recoverable network inputs:} In order to solve the user variables in terms of the network variables (as was described in Section~\ref{sec:motivating-example}), the algorithm is parametrised by an abstract algebraic solver that is used to invert equalities. 
    The power of the solver determines which transformations of user variables can be used when constructing network inputs. For instance, instantiating the algorithm with a solver limited to linear arithmetic results in support for only linear transformations, whereas support for solving quadratic equations permits quadratic transformations. More powerful solvers (e.g. general-purpose symbolic algebra systems) permit a wider class of polynomial and other non-linear transformations.
\end{enumerate}
\section{Background}
\label{sec:background}

Throughout this paper, we assume that the neural networks operate over some numeric domain $\numType$ (e.g. $\real$, $\rationals$, IEEE-754 floats) and that we can write tensor literals over $\numType$ using standard tensor notation (e.g. \texttt{[[0.5, 0.2],[0.1, 0.5]]}).

\subsection{Source language}

Figure~\ref{fig:vehicle-syntax} contains a typed calculus for writing high-level specifications. It includes operations for logic and size-safe tensor operations such as pointwise addition and multiplication, indexing into tensors \texttt{x[i]} and stacking of sub-tensors {$\texttt{[e1, ..., en]}$}. Identifiers for neural networks, represented as $f$ in the grammar, are drawn from a finite, pre-declared set of neural networks $\hNetCtx$. In the typing rules $\Gamma$ is a mapping from the quantified variables to their types.

\begin{figure}[tbp]
\input{figures/syntax-high-level}
\vspace{-1em}
\par\noindent\rule{\textwidth}{0.4pt}
\input{figures/types-high-level}
\caption{Syntax and type-system for the high-level neural network specification language. The standard typing rules for variables ($x$), Boolean operations (\andSymbol{}, \orSymbol{}, \notSymbol{}), Boolean literals ($b$), and tensor literals ($t$) are omitted for brevity.}
\label{fig:vehicle-syntax}
\end{figure}
We emphasise this is a \emph{core} language. A usable neural-network specification language should contain features such as function abstraction, polymorphism and the ability to declare networks and external datasets.
However, we assume these features are dealt with by prior compiler passes and therefore are not included in the language.

\subsection{Target language}

VNN-LIB is the standard query format for neural networks solvers~\cite{vnnlib2025}. Figure~\ref{fig:vnnlib-syntax} shows the syntax for the fragment of VNN-LIB 2.0 we target in this paper.
A VNN-LIB query consists of a list of typed network declarations followed by a list of assertions. The semantics of a VNN-LIB query is ``\emph{given network models that match the network declarations, does there exist an assignment to the network inputs that satisfies all assertions simultaneously?}''. We refer readers to the VNN-LIB standard~\cite{vnnlib2025} for a detailed explanation of the syntax.

\begin{figure}[tbp]
\input{figures/syntax-vnnlib}
\vspace{0.5em}
\par\noindent\rule{\textwidth}{0.4pt}
\vspace{-1.5em}
\input{figures/syntax-query-tree}
\caption{The target language. Above the line is the fragment of the VNN-LIB~2.0 grammar~\cite{vnnlib2025} targeted in this paper (see Figure~\ref{fig:vehicle-syntax} for definition of \texttt{<comp>}). Below the line is a simple tree structure for combining multiple queries. Superscript $^*$ indicates zero or more, $^+$ indicates one or more, $^?$ indicates zero or one.}
\label{fig:vnnlib-syntax}
\end{figure}

In the general case, a high-level specification may require being compiled multiple VNN-LIB queries. Therefore the final target language is the \texttt{<queryTree>} construct that consists of a tree of conjunctions and disjunctions. The leaves of the tree can contain a Boolean literal $b$ or a (possibly negated) list of implicitly disjuncted VNN-LIB queries.

\section{Algorithm}
\label{sec:correct-algorithm}

\begin{figure}[pt]
    \centering
    \input{figures/syntaxprotoquery}
    \vspace{1em}
    
    \newcommand{\varName}{u}

\begin{tikzpicture}[
  every node/.style={font=\small},
  level 1/.style={level distance=0.8cm, sibling distance=3.2cm},
  level 2/.style={level distance=0.8cm, sibling distance=1.6cm}
]

% --- Tree ---
\node (v) at (0,0) {$\varName$}
  child { node {$\varName_0$}
    child { node {$\varName_{00}$} }
    child { node {$\varName_{01}$} }
  }
  child { node {$\varName_1$}
    child { node {$\varName_{10}$} }
    child { node {$\varName_{11}$} }
  };
  
% --- Matrix ---
\node (matrix) at (5.5,-1) {
    $\begin{array}{ll}    
    \varName 
    &= [\varName_0, \varName_1]
    \\
    &= [\varName_0, [\varName_{10},\varName_{11}]]
    \\
    &= [[\varName_{00},\varName_{01}], \varName_1]
    \\
    &= [[\varName_{00},\varName_{01}], [\varName_{10},\varName_{11}]]
    \end{array}$
};
\end{tikzpicture}
    \caption{Intermediate language. \textbf{Top}: the syntax for the language (see Figure~\ref{fig:vehicle-syntax} for definition of classes $\texttt{<comp>}$). \textbf{Bottom left}: the set of hierarchically structured variables generated by a single $2 \times 2$ tensor variable $u$. \textbf{Bottom right}: the various ways variable $u$ can be represented using these variables.}
    \label{fig:intermediate-syntax}
\end{figure}

We now describe the algorithm for compiling an expression in the specification language in Figure~\ref{fig:vehicle-syntax} to an equisatisfiable tree of queries in Figure~\ref{fig:vnnlib-syntax}. 
Compilation proceeds via the intermediate language shown in Figure~\ref{fig:intermediate-syntax}, which is optimised for the elimination of user variables. 
We will refer to a top-level term in the intermediate language as a list of implicitly disjuncted \emph{proto-queries}. 
Each proto-query represents a satisfaction problem which has a consistent solution for the user variables in terms of the network variables. As described in Algorithm~\ref{alg:eliminate-variable}, a proto-query may therefore end up being refined into multiple further proto-queries when eliminating a user variable, as different parts of the tree of the constraints may require different solutions.

As can be seen in Figure~\ref{fig:example-uneliminated-protoquery}, unlike expressions in the specification language, there are no quantifiers, negations, network applications, stacking or indexing operations in proto-queries. However, unlike VNN-LIB, the proto-queries still have arithmetic operations over tensors. 
As will be described by Algorithm~\ref{alg:compile-linear-expr}, elimination of network applications will be achieved by replacing them with fresh network variables representing their input and output. The elimination of the stacking and indexing operations will be achieved by associating every tensor variable (both user variables and network variables) with a hierarchy of child variables that represent every possible sub-index of that tensor (see Figure~\ref{fig:intermediate-syntax}).

\subsection{Compiling a query tree}

\begin{figure}[p]
\centering
    \begin{subfigure}{\textwidth}
        \begin{tikzpicture}
    % Top left 
\draw [draw=white] (-5,3.7) rectangle (6.5,2) node[midway, black, align=left] {
\texttt{\existsSymbol{} (u v w : \hRealTensorType{[2,2]}). } \\
    $\qquad$ \texttt{\hLeq{0}{f(u)}\,\andSymbol{}\,\hLeq{f(v)}{1}\,\andSymbol{}\,\hLeq{0}{f([w[0],w[1]])}} 
    \,\andSymbol{}
    \\ 
    $\qquad$ \texttt{\hLeq{f([w[0],[u[1][0],v[1][1]]])}{1}\,\andSymbol{}} \\
    $\qquad$ \texttt{\hEq{u}{v}\,and\,\hEq{u[0]}{w[0]}}
};
\end{tikzpicture}
        \caption{Initial high-level specification. Network $f$ has type $\numType^{2\times2} \rightarrow \numType$.}
        \label{fig:example-specification}
    \end{subfigure}
    \vspace{0.1em}

    \begin{subfigure}{\textwidth}
        \begin{tikzpicture}
    % Top left 
\draw [draw=white] (-5,3.7) rectangle (6.5,2) node[midway, black, align=left] {
\texttt{\hEq{$x^1$}{u}\,\andSymbol{}\,\hEq{$x^2$}{v}\,\andSymbol{}\,\hEq{$x^3_0$}{w$_0$}\,\andSymbol{}\,\hEq{$x^3_1$}{w$_1$}\,\andSymbol{}}
\\
\texttt{\hEq{$x^4_0$}{w$_0$}\,\andSymbol{}\,\hEq{$x^4_{10}$}{u$_{10}$}\,\andSymbol{}\,\hEq{$x^4_{11}$}{v$_{11}$}}\,\andSymbol{}\\
\texttt{\hAnd{\hLeq{0}{$y^1$}}
{\hLeq{$y^2$}{1}}\,\andSymbol{}\,\hAnd{\hLeq{0}{$y^3$}}{\hLeq{$y^4$}{1}}\,\andSymbol{}}
\\ 
    \texttt{\hEq{u}{v}\,and\,\hEq{u$_0$}{w$_0$}}
};
\end{tikzpicture}
        \caption{Proto-query generated before the first call to Algorithm~\ref{alg:eliminate-variable} to eliminate variable $w$.}
        \label{fig:example-uneliminated-protoquery}
    \end{subfigure}
    \vspace{0.1em}

    \begin{subfigure}{\textwidth}
        \begin{tikzpicture}
\draw [draw=white] (-5,3.3) rectangle (6.5,2) node[midway, black, align=left] {
\texttt{\hEq{$x^4_{0}$}{$x^3_{0}$}\,and\,\hEq{$x^4_{10}$}{$x^1_{10}$}\,and\,\hEq{$x^4_{10}$}{$x^2_{10}$}\,and\,
}
\\
\texttt{\hAnd{\hLeq{0}{$y^1$}}
{\hLeq{$y^2$}{1}}\,\andSymbol{}\,\hAnd{\hLeq{0}{$y^3$}}{\hLeq{$y^4$}{1}}\,\andSymbol{}}
\\ 
    \texttt{\hEq{$x^1$}{$x^2$}\,and\,\hEq{$x^1_0$}{$x^3_0$}
    }
};
\end{tikzpicture}
        \caption{Proto-query after eliminating all three user variables via the call to Algorithm~\ref{alg:elim-exists}.}
        \label{fig:example-eliminated-protoquery}
    \end{subfigure}
    \vspace{0.1em}

    \begin{subfigure}{\textwidth}
        \begin{tikzpicture}[
  every node/.style={font=\small}
]

% --- Tree ---
\newcommand{\graphnodes}[4]{
    \node (a#1) at (#2-0,#3-0) {$x^1_{#1}$};
    \node (b#1) at (#2+1,#3-0) {$x^2_{#1}$};
    \node (c#1) at (#2-0,#3-1) {$x^3_{#1}$};
    \node (d#1) at (#2+1,#3-1) {$x^4_{#1}$};
    \node [minimum width=2cm,draw,circle,#4] (x#1) at (#2+0.5,#3-0.5) {};
    \coordinate (bottom#1) at (#2+0.5,#3-2.5);
};

\tikzset{
    explicitEq/.style={},
    inheritEq/.style={dashed},
    downward/.style={},
    downwardSet/.style={shift={(0,0.5)},fill=white},
    upwardSet/.style={fill=white}
}

\tikzset{
    largearrow/.style={
        -{Stealth[length=2mm, width=2mm]} % Customize length and width
    }
}

\newcommand{\dx1}{2}
\newcommand{\dx2}{2}
\newcommand{\y1}{3}
\newcommand{\y2}{3}

% Real spec
\draw [draw=white] (-5.4,0.7) rectangle (6.7,-10);

\draw [explicitEq] (2.2,-0.2) -- (3,-0.2);
\node [anchor=west,align=left] at (3.2,-0.2) {Explicit equality};

\draw [inheritEq] (2.2,-0.8) -- (3,-0.8);
\node [anchor=west,align=left] at (3.2,-0.8) {Inherited equality};

\graphnodes{}{0}{0}{}
\coordinate (top) at (0.5,1.5);
\draw [explicitEq] (a) -- (b);

\graphnodes{0}{-3}{-2}{}
\draw [inheritEq] (a0) -- (b0);
\draw [explicitEq] (a0) -- (c0);
\draw [explicitEq] (c0) -- (d0);

\graphnodes{1}{3}{-2}{}
\draw [inheritEq] (a1) -- (b1);

\graphnodes{00}{-4.5}{-4.8}{dotted}
%\draw [inheritEq] (a00) -- (b00);
%\draw [explicitEq] (c00) -- (d00);
%\draw [inheritEq] (a00) -- (c00);

\graphnodes{01}{-1.5}{-4.8}{dotted}
%\draw [inheritEq] (a01) -- (b01);
%\draw [explicitEq] (c01) -- (d01);
%\draw [inheritEq] (a01) -- (c01);

\graphnodes{10}{1.5}{-4.8}{}
\draw [inheritEq] (a10) -- (b10);
\draw [explicitEq] (a10) -- (d10);

\graphnodes{11}{4.5}{-4.8}{}
\draw [inheritEq] (a11) -- (b11);
\draw [explicitEq] (b11) -- (d11);

% Downwards
\node (initial) [downwardSet] at ([xshift=-3.5cm,yshift=-0.5cm]x) {$\{x^1\},\{x^2\},\{x^3\},\{x^4\}$};
\draw [downward,largearrow] (initial) -- (x);

\draw [downward] (x) -- (bottom);
\draw [downward,largearrow] (bottom) -- (x0);
\draw [downward,largearrow] (bottom) -- (x1);
\node [downwardSet] at (bottom) {$\{x^1,x^2\},\{x^3\},\{x^4\}$};

% \draw (x0) -- ([yshift=-0.8cm]bottom0);
% \draw [downward,largearrow] ([yshift=-0.8cm]bottom0) -- (x00);
% \draw [downward,largearrow] ([yshift=-0.8cm]bottom0) -- (x01);
\node (res0) [fill=white] at  (-2.5,-8.5) {$\{x^1,x^2,x^3,x^4\}$};
\draw [largearrow] (x0) -- (res0);
\node [downwardSet] at (bottom0) {$\{x^1,x^2,x^3,x^4\}$};

\draw (x1) -- ([yshift=-0.8cm]bottom1);
\draw [downward,largearrow] ([yshift=-0.8cm]bottom1) -- (x10);
\draw [downward,largearrow] ([yshift=-0.8cm]bottom1) -- (x11);
\node [downwardSet] at (bottom1) {$\{x^1,x^2\},\{x^3\},\{x^4\}$};

% Upwards

% Lowest points
% \node (res00) [upwardSet] at (bottom00) {$\{x^1,x^2,x^3,x^4\}$};
% \draw [largearrow] (x00) -- (res00);

% \node (res01) [upwardSet] at (bottom01) {$\{x^1,x^2,x^3,x^4\}$};
% \draw [largearrow] (x01) -- (res01);

\node (res10) [upwardSet] at (bottom10) {$\{x^1,x^2,x^4\},\{x^3\}$};
\draw [largearrow] (x10) -- (res10);

\node (res11) [upwardSet] at (bottom11) {$\{x^1,x^2,x^4\},\{x^3\}$};
\draw [largearrow] (x11) -- (res11);

% Middle layer
%\draw [largearrow] (res01) -| (res0);

\node (res1) [upwardSet] at (3.5,-8.5) {$\{x^1,x^2,x^4\},\{x^3\}$};
\draw [largearrow] (res10) -| (res1);
\draw [largearrow] (res11) -| (res1);

% Bottom layer

\node (res) [upwardSet] at (0.5,-9.5) {$\{x^1,x^2,x^4\},\{x^3\}$};
\draw [largearrow] (res0) -| (res);
\draw [largearrow] (res1) -| (res);
\end{tikzpicture}
        \caption{Operation of Algorithm~\ref{alg:equality-elim} on the proto-query in Figure~\ref{fig:example-eliminated-protoquery} to optimise the number of applications of $f$ by computing the equivalence classes induced by the hierarchical multi-graph. Each circle represents a graph of the equalities at that tensor subindex. The final result indicates we can replace $(x^2, y^2)$ and $(x^4, y^4)$ with $(x^1, y^1)$.}
        \label{fig:example-equality-elimination}
    \end{subfigure}

    \caption{Compilation of an example specification (continued next page).}
    \label{fig:example}
\end{figure}

\begin{figure}\ContinuedFloat
    
    \begin{subfigure}{\textwidth}
        \begin{tikzpicture}
\draw [draw=white] (-5,3) rectangle (6.5,2) node[midway, black, align=left] {
\hAnd{\hLeq{0}{$y^1$}}
{\hLeq{$y^1$}{1}} 
\andSymbol{} 
\hAnd{\hLeq{0}{$y^3$}}{\hLeq{$y^1$}{1}} 
\andSymbol{}
\\
    \hEq{$x^1_0$}{$x^3_0$} \andSymbol{} \hEq{$x^1_{00}$}{$x^3_{00}$}
    \andSymbol{} \hEq{$x^1_{01}$}{$x^3_{01}$}
    };
\end{tikzpicture}
        \caption{The proto-query after using the result of Figure~\ref{fig:example-equality-elimination} to eliminate the redundant network applications in Figure~\ref{fig:example-eliminated-protoquery}.}
        \label{fig:example-optimised}
    \end{subfigure}
    \vspace{0.5em}
    
    \begin{subfigure}{\textwidth}
        \input{figures/example-vnnlib}
        \caption{Final VNN-LIB query generated from the proto-query in Figure~\ref{fig:example-optimised}.}
        \label{fig:example-vnnlib-query}
    \end{subfigure}
    
\caption{Continued from previous page.}
\end{figure}

The specification shown in Figure~\ref{fig:example-specification} will serve as a running example throughout our presentation of the algorithm.
It is deliberately constructed to illustrate a wide range of complications that may arise in practice.
For more natural specifications that nonetheless exhibit many of these challenges, we refer the reader to the examples discussed in Section~\ref{sec:evaluation}.

\begin{algorithm}[p]
\caption{Entry point for compiling a  specification}
\label{alg:compile}
\begin{algorithmic}[1]
\Require $e$ is an expression typable as $\hNetCtx, \varnothing \vdash e : \hBoolType$. 
\Ensure Returns a VNN-LIB query tree which is equisatisfiable with $e$.
\Function{\compileSymbol}{$e$}
    \Switch{$e$}
        \Case{$b$}
        \Return QueryTreeTrivialLeaf $b$
        \EndCase        
        \Case{$\hNot{e}$}
            %\State 
            \Return \compile{\elimNot{$e$}}
        \EndCase
        \Case{$\hOr{e_1}{e_2}$}
            %\State
            \Return QueryTreeOr(\compile{$e_1$}, \compile{$e_2$})
        \EndCase
        \Case{$\hAnd{e_1}{e_2}$}
            %\State 
            \Return QueryTreeAnd(\compile{$e_1$}, \compile{$e_2$})
        \EndCase
        % \Case{$\hIf{e_1}{e_2}{e_3}$}
        %     %\State
        %     \Return \compile{($e_1$ and $e_2$) or (not $e_1$ and $e_3$)}
        % \EndCase
        \Case{$e_1$ op $e_2$}
        \Return QueryTreeTrivialLeaf (\evalTrivialComparison{op}{$e_1$}{$e_2$})
        \EndCase
        \Case{$\hExists{v}{\tau}{e}$}
            \State $protoqueries \gets \elimExists{v}{e}$
            \State $queries \gets \optimiseQuery{protoqueries}$            
            \State \Return QueryTreeLeaf queries
        \EndCase
        \Case{$\hForall{v}{\tau}{e}$}
            \State $protoqueries \gets \elimExists{v}{\hNot{e}}$
            \State $queries \gets \optimiseQuery{protoqueries}$
            \State \Return QueryTreeNegatedLeaf queries
        \EndCase
    \EndSwitch
\EndFunction
\end{algorithmic}
\vspace{-2em}
\end{algorithm}

\paragraph{Algorithm \ref{alg:compile}: \compileSymbol{}}  is the entry point for the compiler. It proceeds by pattern matching on the boolean structure of the specification. Trivial boolean values are returned, and `\notSymbol{}' constructors are eliminated via \elimNotSymbol{}, which takes an expression $e$ and pushes a \notSymbol{} through the expression down to the level of comparisons via De Morgan's laws, e.g. $\elimNot{\hAnd{x >= 7}{y < 3}}$ evaluates to $\hOr{x < 7}{y >= 3}$.
The `\andSymbol{}' and `\orSymbol{}' constructors are preserved. Comparisons are evaluated using  \evalTrivialComparisonSymbol{} as there can be no free variables as no quantifiers have been encountered yet.
When an `\existsSymbol{}' is encountered, first it is eliminated by compiling it to a set of proto-queries via Algorithm~\ref{alg:elim-exists} and then those proto-queries are compiled to VNN-LIB queries via Algorithm~\ref{alg:compile-vnnlib-queries}. When a `\forallSymbol{}' is encountered, it is negated to turn it into a satisfaction problem and then compiled as if it was an `\existsSymbol{}'. In the case of the example specification in Figure~\ref{fig:example-specification}, there is no such higher structure so it simply enters the '\existsSymbol{}' case.

\subsection{Eliminating user variables}
\label{sec:properties}

\begin{algorithm}[t]
\caption{Eliminating an existential quantifier}
\label{alg:elim-exists}
\begin{algorithmic}[1]
\Require (i) $var$ is a user variable typable as $\hNetCtx,\Gamma \vdash var : \hRealTensorType{ds}$, 
\Statex \quad\quad\quad (ii) $body$ is an expression typable as $\hNetCtx,\Gamma[var \rightarrow \hRealTensorType{ds}] \vdash body : \hBoolType{}$.
\Ensure Outputs implicitly disjuncted proto-queries which do not refer to $var$.
\Function{\elimExistsSymbol}{$var$, $body$}
    \State $protoqueries \gets \compileBodyRec{body}$
    \State \Return $\elimVar{var}{finalProtoqueries}$ \label{algl:eliminate-user-variable} 
\EndFunction
\end{algorithmic}
\vspace{-2em}
\end{algorithm}

\paragraph{Algorithm~\ref{alg:elim-exists}: \elimExistsSymbol{}} eliminates an existentially quantified user variable from the specification, and is performed in two stages: first, the quantifier body is recursively compiled into a set of proto-queries, and second, the user variable is eliminated from those proto-queries.

\begin{algorithm}[t]
\caption{Compiling the body and equalities of an existential quantifier}
\label{alg:compile-body-rec}
\begin{algorithmic}[1]
\Require $expr$ is a an expression typable as $\hNetCtx,\Gamma \vdash expr : \hBoolType{}$, 
\Ensure Outputs implicitly disjuncted proto-queries which do not refer to $var$.
\Function{\compileBodyRecSymbol{}}{$expr$}
\State $protoqueries \gets \compileBody{expr}$
\If{no network equalities noted during compilation of expr}
\State \Return $protoqueries$
\Else
\State $equalities \gets \text{conjunction of all noted network equalities}$
\State $equalityProtoqueries \gets \compileBodyRec{equalities}$
\State \Return \andProtoqueries{$protoqueries$}{$equalityProtoqueries$}
\EndIf{}
\EndFunction
\end{algorithmic}
\vspace{-2em}
\end{algorithm}

\paragraph{Algorithm~\ref{alg:compile-body-rec}}: \compileBodyRecSymbol{} compiles the constraints in the body of the existential quantifier.
During this process (in Algorithm~\ref{alg:compile-linear-expr}), network applications may be encountered, which generate additional equality constraints.
For example, the application \texttt{f(u)} in Figure~\ref{fig:example-specification} generates the equality \texttt{\hEq{$x^1$}{u}} shown in Figure~\ref{fig:example-uneliminated-protoquery}.
These equalities are themselves compiled and conjuncted with the resulting proto-queries.

\begin{algorithm}[t]
\caption{Compiling the body of an existential quantifier}
\label{alg:compile-body}
\begin{algorithmic}[1]
\Require $e$ is a normalised expression of type $\hBoolType$. 
\Ensure Outputs a list of implicitly disjuncted proto-queries.
\Function{compileBody}{$e$}
    \Switch{$e$}
        \Case{$b$}
            \Return $b$
        \EndCase
        \Case{$\hNot{e}$}
            %\State 
            \Return \compileBody{\elimNot{$e$}}
        \EndCase
        \Case{$\hOr{e_1}{e_2}$}
            %\State
            \Return \orProtoqueries{\compileBody{$e_1$}}{\compileBody{$e_2$}}
        \EndCase
        \Case{$\hAnd{e_1}{e_2}$}
            %\State 
            \Return \andProtoqueries{\compileBody{$e_1$}}{\compileBody{$e_2$}}
        \EndCase
        % \Case{$\hIf{e_1}{e_2}{e_3}$}
        %     \Return \compileBody{($e_1$ and $e_2$) or (not $e_1$ and $e_3$)}
        % \EndCase
        \Case{$e_1$ op $e_2$}
            \Return \compileComparison{op}{$e_1$}{$e_2$}
        \EndCase
        \Case{$\hExists{v}{\tau}{e}$}
        	\Return $\elimExists{v}{e}$
        \EndCase
        \Case{$\hForall{v}{\tau}{e}$}
            \Throw AlternatingQuantifiersError
        \EndCase
    \EndSwitch
\EndFunction
\end{algorithmic}
\vspace{-2em}
\end{algorithm}

\paragraph{Algorithm~\ref{alg:compile-body}: \compileBodySymbol{}} describes how the body of the existential quantifier may be compiled to a list of implicitly disjuncted proto-queries.
Trivial booleans are immediately returned and `\notSymbol{}' constructors are again eliminated by pushing them inwards. When encountering `\orSymbol{}' and `\andSymbol{}', their arguments are compiled recursively and the resulting proto-queries combined via \orProtoqueriesSymbol{} and \andProtoqueriesSymbol{} (see Appendix~\ref{app:proto-query-logic} for discussion of implementation).
When a binary comparison is encountered, it is compiled using Algorithm~\ref{alg:compile-comparison}.
When an `\existsSymbol{}' is encountered we can recursively call Algorithm~\ref{alg:elim-exists} to eliminate it. However, if a `\forallSymbol{}' is encountered, we know the specification contains alternating quantifiers and therefore are forced to error (see Section~\ref{sec:limitations} for discussion)\footnote{See~\cite{DaggittAKKA23} for how meaningful explanations of this error can be generated for the user.}.

\begin{algorithm}[tbp]
\caption{Compiling a high-level comparison}
\label{alg:compile-comparison}
\begin{algorithmic}[1]
\Require $e_1$ and $e_2$ are expressions typable as $\hNetCtx, \Gamma \vdash e_1, e_2 : \hRealTensorType{ds}$. 
\Ensure A list of implicitly disjuncted queries equisatisfiable with $\bop{e_1}{op}{e_2}$ when conjuncted with the set of generated network equalities.
\Function{compileComparison}{$e_1$, op, $e_2$}
\try
    \State $e_1' \gets \purify{e_1}{0}$
    \State $e_2' \gets \purify{e_2}{0}$
    \State \Return [$e_1'$ op $e_2'$]
\Catch{FoundStack $d$}
    \If{op = `!='}
    \State $e' \gets \bigvee_{i=0}^d~\bop{(\evalAt{e_1}{i})}{op}{(\evalAt{e_2}{i})}$
    \Else
    \State $e' \gets \bigwedge_{i=0}^d~\bop{(\evalAt{e_1}{i})}{op}{(\evalAt{e_2}{i})}$
    \EndIf
    \State \Return \compileBody{$e'$}
\EndTry
\EndFunction
\end{algorithmic}
\vspace{-2em}
\end{algorithm}

\paragraph{Algorithm~\ref{alg:compile-comparison}: \compileComparisonSymbol{}} uses Algorithm~\ref{alg:compile-linear-expr} to try to compile the LHS and RHS of the comparison to arithmetic expressions in the intermediate proto-query language. 
Recall that arithmetic proto-query expressions must consist only of arithmetic operations over variables and cannot contain tensor stacking and indexing operations. As the description of Algorithm~\ref{alg:compile-linear-expr} will explain, compilation of an arithmetic expression will fail if it contains if it contains a stacking operation. In this case the pointwise comparison is rewritten as a set of conjunctions/disjunctions of comparisons of one lower dimension and recursively compiled. For example, the application in \texttt{f([w[0],[u[1][0],v[1][1]]])} in Figure~\ref{fig:example-specification} generates the constraint  \texttt{\hEq{$x^4$}{[w[0],[u[1][0],v[1][1]]]}}). This cannot be compiled to a single equality so is recursively broken down into the equalities \texttt{\hEq{$x^4_0$}{w$_0$}\,\andSymbol{}\,\hEq{$x^4_{10}$}{u$_{10}$}\,\andSymbol{}\,\hEq{$x^4_{11}$}{v$_{11}$}} in Figure~\ref{fig:example-uneliminated-protoquery}.

\begin{algorithm}[t]
\caption{Compiling an arithmetic expression}
\label{alg:compile-linear-expr}
\begin{algorithmic}[1]
\newcommand{\dims}{\ensuremath{ds}}

\Require $e$ is a value of type $\hRealTensorType{ds}$ for some $ds$ such that $|ds| \leq n$. 
\Ensure Outputs an equivalent arithmetic expression (modulo the network input equalities generated).
\Function{\purifySymbol{}}{$e$, $n$}
    \Switch{($e$, $n$)}
        \Case{($t$, _)}
            \Return $t$
        \EndCase
        \Case{($\hAdd{e_1}{e_2}$, _)}
            \Return \purify{$e_1$}{$n$} + \purify{$e_2$}{$n$}
        \EndCase
        \Case{($\hMul{e_1}{e_2}$, _)}
            \Return \purify{$e_1$}{$n$} * \purify{$e_2$}{$n$}
        \EndCase
        \Case{($f e$, _)}
            \State x, y $\gets$ fresh network variables
            \State \Note{\hEq{x}{e}} \label{line:note-equality}
            \State \Return \purify{y}{$n$}
        \EndCase
        \Case{($\hAt{e}{i}$, _)}
            \State $e' \gets$ \evalAt{\purify{$e$}{$n+1$}}{$i$}
            \State \Return \purify{$e'$}{$n$}
        \EndCase
        \Case{($\hSeq{e_1 \ldots e_d}$, 0)}
            \Throw{FoundStack $d$}
        \EndCase
        \Case{($\hSeq{e_1 \ldots e_d}$, $m + 1$)}
            \Return $\hSeq{e_1 \ldots e_d}$
        \EndCase
        \Case{($x$, 0)}
            \Return $x$
        \EndCase
        \Case{($x$, $m+1$)}
            \Return $\hSeq{x_0, \ldots , x_n}$
        \EndCase
        % \Case{($\hIf{e_1}{e_2}{e_3}$, _)}
        %     \Throw{FoundIf}
        % \EndCase
    \EndSwitch
\EndFunction
\end{algorithmic}
\vspace{-2em}
\end{algorithm}

\paragraph{Algorithm~\ref{alg:compile-linear-expr}: \purifySymbol{}} compiles an arithmetic expressions $e$ to the intermediate language, where the second parameter $n$ tracks how many extra dimensions the current expression has than the original arithmetic expression being compiled. When a tensor literal is encountered, it is returned immediately, and addition and multiplication are recursed upon in the expected way. If a network application is encountered then two fresh variables~$x$ and~$y$ are generated to represent the inputs and outputs respectively. We note that~$x$ must be equal to the input expression~$e$ ready to be picked up by Algorithm~\ref{alg:compile-body}\footnote{We cannot append the equality to the tree here as the result would be ill-typed.}. We then return the result of compiling the output variable~$y$.

The next case is tensor indexing, \texttt{\hAt{e}{i}}. In order to eliminate the indexing operation we first compile $e$, the tensor being indexed into, noting that the current depth has increased. The result is then evaluated using \evalAtSymbol{}, which takes a tensor expression $e$ and index $i$ and evaluates $e$ at index $i$. When evaluation is blocked by variables, \evalAtSymbol{} automatically distributes the indexing operation across the arithmetic expressions, e.g. $\evalAt{x+y}{i} = \evalAt{x}{i} + \evalAt{y}{i}$.
Thanks to this assumption, and taking into account the remaining cases of Algorithm~\ref{alg:compile-linear-expr}, this indexing operation is guaranteed to reduce, with compilation then continuing on the returned sub-tensor. The definition of \evalAtSymbol{} is available in Appendix~\ref{app:eval-lookup}.

If a stack operation is encountered and the current depth is zero, then it is impossible to compile the expression to a proto-query expression and therefore an error is thrown which will be caught by Algorithm~\ref{alg:compile-comparison}. Otherwise, if the current depth is non-zero then the stack operation is returned unchanged as it will eventually be eliminated by a call to \evalAtSymbol{}. 
If a variable at depth 0 is encountered, it is immediately returned. Otherwise, if encountered at a non-zero depth, it is substituted for its stacked child variables (see Figure~\ref{fig:intermediate-syntax}), which again will eventually be eliminated by a call to \evalAtSymbol{}.

We have now completely described how to compile the body of an existential quantifier to a set of proto-queries. Figure~\ref{fig:example-uneliminated-protoquery} shows the state of the proto-query at this point. We will now return to the use of Algorithm~\ref{alg:eliminate-variable} in Algorithm~\ref{alg:elim-exists} that eliminates the quantified user variable from those proto-queries.

\begin{algorithm}[t]
\caption{Eliminating a user variable}
\label{alg:eliminate-variable}
\newcommand{\var}{v}

\begin{algorithmic}[1]
\Require (i) a tensor variable $\var$ to eliminate and (ii) a list of $protoqueries$. 
\Ensure Returns a list of equisatisfiable proto-queries that do not contain references to $\var$ or any of its child variables.
\Function{\elimVarSymbol{}}{$\var$, $protoqueries$}
	\State $result$ $\gets$ $\emptyList{}$
	\For{$protoquery$ in $protoqueries$}
		\State $equalityConstrainedProtoqueries$ $\gets$ \equalitySolver{$\var$}{$protoquery$}
		\For{$childProtoquery$ in $equalityConstrainedProtoqueries$}
			\If{found equality $\var = e$ for some $e$}
				\State $solvedProtoquery \gets$ substitute $\var = e$ through $childProtoquery$.
			 	\State append $solvedProtoquery$ to result 
			\ElsIf{$\var$ has child variables $v_0, v_1, ... v_n$}
				\State $a_0 \gets$ substitute $\var = [v_0, v_1, ..., v_n]$ through $childProtoquery$.
				\State $a_1 \gets \elimVar{a_0}{v_0}$
				\State $a_2 \gets \elimVar{a_1}{v_1}$
				\State ...
				\State $a_{n + 1} \gets \elimVar{a_n}{v_n}$
				\State concatenate $a_{n+1}$ to $result$
            \Else{} \Throw{UninvertableInput}
			\EndIf
		\EndFor
	\EndFor
	\State \Return $result$
\EndFunction
\end{algorithmic}
\vspace{-2em}
\end{algorithm}

\paragraph{Algorithm~\ref{alg:eliminate-variable}: \elimVarSymbol{}} iterates over each proto-query in the list and invoking an abstract \equalitySolverSymbol{} on it to try and find and solve an equality over the current variable. For example, in Figure~\ref{fig:example-uneliminated-protoquery} to Figure~\ref{fig:example-eliminated-protoquery} when trying to eliminate user variable $u$ the equality solver would find $x^1 == u$.
As protoqueries may contain disjunctions internally, different disjunctions may contain different solutions for the user variable, e.g. the results of calling the solver on the proto-query:
\begin{center}
    $\hOr{(\hAnd{v == x}{...})}{\hOr{(\hAnd{v == x + 2}{...})}{...}}$
\end{center}
will return three different proto-queries, one where $v == x$, one where $v == x + 2$ and one where there is no solution for $v$.
Note that the solver must have access to the whole proto-query, not just the network input equalities, as there may be other equalities in the specification that it can use, e.g. in
\begin{center}    
$\hExists{u}{\hRealTensorType{[2]}}{ \hExists{v}{\hRealTensorType{[2]}}{\hAnd{f(u) \leq 1}{u == v}}}$.
\end{center}
variable $v$ can only be eliminated via the equality $u == v$.

For each of the proto-queries returned by the solver, if an equality over the target variable was found, then the equality is substituted through the proto-query immediately. 
This eliminates the target tensor variable without having to solve for each of its elements individually. This is one of the two crucial optimisations that allows efficient compilation in the common case and prevents the blow-up in compilation time discussed in Section~\ref{sec:motivating-example}.

If the solver is unable to find an equality for the target variable in the returned proto-query, the algorithm first checks if the target variable has non-empty dimensions. If so, then the algorithm falls-back to eliminating each of its child variables that represent the rows of that tensor.
In the worst-case, this algorithm therefore performs a depth-first search over the tensor structure of the variable, as shown in Figure~\ref{fig:intermediate-syntax}. 
If the target variable is zero-dimensional, then the algorithm errors as it is unable to solve for the user variable. See Section~\ref{sec:limitations} for a discussion of how the precise implementation of \equalitySolverSymbol{} impacts when this error is thrown.

\subsection{Query optimisation}

At this point, the description of \elimExistsSymbol{} used in Algorithm~\ref{alg:compile} is now complete, and the list of proto-queries is guaranteed to contain only network variables. See Figure~\ref{fig:example-eliminated-protoquery} for the state of the running example. 
The remaining task is to compile these proto-queries to VNN-LIB queries.

\begin{algorithm}[ht]
\caption{Compiling protoqueries to VNN-LIB}
\label{alg:compile-vnnlib-queries}
\begin{algorithmic}[1]
\Require a list of $protoqueries$ with only network variables
\Ensure output is an equisatisfiable list of VNN-LIB queries.
\Function{\optimiseQueriesSymbol{}}{$protoqueries$}
    \State $result \gets \emptyList{}$
    \For{$protoquery$ in $protoqueries$}
        \State $optimisedProtoquery \gets \eliminateEqualApps{protoquery}$
        \State $reducedProtoquery \gets \reduceDims{optimisedProtoquery}$
        \State $vnnlibQuery \gets \emptyset$
        \State append all network applications in $reducedProtoquery$ to $vnnlibQuery$
        \State append all constraints in $reducedProtoquery$ to $vnnlibQuery$
        \State append $vnnlibQuery$ to $result$
    \EndFor
    \State \Return $result$
\EndFunction
\end{algorithmic}
\vspace{-2em}
\end{algorithm}

\paragraph{Algorithm~\ref{alg:compile-vnnlib-queries}: \optimiseQueriesSymbol{}} compiles each proto-query into a corresponding VNN-LIB query in three steps. Firstly, the number of network applications in the proto-query will be optimised via Algorithm~\ref{alg:equality-elim}. Next, the call to \reduceDimsSymbol{} reduces all multi-dimensional tensor comparisons to a set of conjuncted/disjuncted comparisons over zero-dimensional tensors (definition not given as it uses the same method as in \textbf{\textsc{catch}} clause of Algorithm~\ref{alg:compile-comparison}). 
Finally, the resulting proto-query can be trivially compiled to VNN-LIB.

\paragraph{Algorithm~\ref{alg:equality-elim}: \eliminateEqualAppsSymbol{}} attempts to eliminate redundant network applications introduced by Algorithm~\ref{alg:elim-exists}.
Consider the specification:
\begin{lstlisting}{haskell}
exists (a : Tensor []). f a > 0 and f a < 1
\end{lstlisting}
In this case Algorithm~\ref{alg:elim-exists} will introduce two network applications, when clearly only one network application is needed. 
This must be avoided as additional network applications result in the solver having to reason about multiple copies of the same network and potentially increases the solve time exponentially.

In order to eliminate redundant network applications it is necessary to detect when the inputs for two applications of the same network are equal.
One approach is to attempt to detect equal inputs when the fresh network variables are generated in Algorithm~\ref{alg:compile-linear-expr}. However, this does not work in general, as two network inputs being equal is a semantic concept rather than a syntactic one. See the example specification in Figure~\ref{fig:example-specification}
where \texttt{f(u)} and \texttt{f(v)} are only provably equal because of the additional equality \texttt{\hEq{u}{v}} in the specification.

Therefore the most natural point to perform duplicate network application elimination is after elimination of the user variables.
Duplicate identification can be done by first computing the equivalence closure over the graph induced by equality constraints present in the proto-query. Finding duplicate network applications is equivalent to finding the connected components in that graph. However, the general problem is complicated by the fact that equalities between input variables may be provable from equalities over their sub-indices, i.e. two tensors are equal if it all their rows are provably equal. 

\begin{algorithm}[h]
    \begin{algorithmic}[1]
\Require a $protoquery$ with only network variables
\Ensure an equisatisfiable $protoquery$ with possibly fewer network applications.
\Function{\eliminateEqualAppsSymbol{}}{$protoquery$}
\State $optimised \gets protoquery$
\For{each $network$ used in $protoquery$}
\State $initialClasses \gets$ every application of $network$ in $protoquery$
\State $initialDims \gets$ input dimensions of network
\State $classes \gets \findConnectedComponents{optimised}{initialClasses}{initialDims}{\emptyList}$
\For{each $class$ in $classes$}
    \If{size $class$ > 1}
        \State $optimised \gets$ eliminate redundant variables in $class$ in  $optimised$
    \EndIf
\EndFor
\EndFor
\State \Return $optimised$
\EndFunction
\State

\Function{\findConnectedComponentsSymbol{}}{$query$, $classes$, $dims$, $is$}
    \If{size of $classes$ = 1} \Return $classes$
    \Else
        \State $\begin{aligned}[t]
            newClasses \gets \,
                & \text{equivalence classes induced by $query$ at indices $is$} \\
                & \text{taking into account equivalences inherited from $classes$}
        \end{aligned}$
        \Switch{$dims$}
            \Case{$\emptyList$}
                \Return $newClasses$
            \EndCase
            \Case{$\cons{d}{ds}$}
                \State $childClasses \gets \bigcup_{i=0}^d \findConnectedComponents{query}{newClasses}{ds}{\cons{i}{is}}$
                \State \Return intersection of $childClasses$
            \EndCase
        \EndSwitch
        
    \EndIf
\EndFunction
\end{algorithmic}
    \caption{Optimising the number of network eliminations}
    \label{alg:equality-elim}
\end{algorithm}

Therefore~\findConnectedComponentsSymbol{} in Algorithm~\ref{alg:equality-elim} recursively computes the equivalence classes over the hierarchy of tensor indices for a particular network. It passes down the connected components found so far and then merges together the resulting equivalence classes on the upwards pass. See Figure~\ref{fig:example-equality-elimination} for a full worked example of this algorithm applied to the proto-query in Figure~\ref{fig:example-eliminated-protoquery} to generate the optimised query in Figure~\ref{fig:example-optimised}.
Compilation is now complete and the final resulting VNN-LIB query can be seen in Figure~\ref{fig:example-vnnlib-query}.
\section{Soundness}

The only non-local transformation of the abstract syntax tree (AST) occurs in the interaction of Algorithms~\ref{alg:compile-body-rec} and~\ref{alg:compile-linear-expr}, where equality constraints generated by the latter are incorporated into the proto-queries constructed by the former. This is sound because each such equality only constrains a freshly introduced network input variable, and therefore lifting them does not alter the semantics of existing variables or subterms.
All remaining steps of the algorithm consist of local transformations of the AST whose soundness can be verified independently, and once verified, the overall soundness of the algorithm follows by composition.

Numerical soundness is an immediate by observing that only $\equalitySolverSymbol$ performs any arithmetic manipulation. Therefore if the solver is numerically sound over type $\numType$ then so is the overall algorithm.

In order to obtain a fully formal proof of soundness, the algorithm would need to be explicitly parameterised by a network theory $\psi$, as defined by the VNN-LIB standard~\cite{vnnlib2025}. One would then use $\psi$ to derive the semantics of VNN-LIB, define the semantics of the high-level specification language and then prove compilation is a semantically preserving operation. Establishing this result would allow the extension of neural network verification certificates~\cite{desmartin2023towards,elboher2025abstraction} to high-level specifications. We leave this formalisation for future work.

\section{Performance}
\label{sec:evaluation}

As with any compilation algorithm, the time complexity depends primarily on the semantics of the input specification rather than its syntactic length. Nevertheless, the following analysis shows that in the worst case the runtime grows exponentially in the number of dimensions of the input tensor.

Let $s$ be the shape of a user variable $u$. The compilation process generates one variable for each sub-index of $u$, yielding a total of $\sum_i \prod_{j = 0}^i s_i$ variables (i.e. $u$, $u_0$, $u_1$, $\ldots$). 
Since all constraints are well-typed, any single assertion can involve at most $\prod_{i} s_i$ variables. 
Consequently, even a trivial implementation of \equalitySolverSymbol{} which finds and rearranges linear equalities over $u$, requires $O(\prod_{i} s_i)$ time if the equalities are dense.
In the worst-case of Algorithm~\ref{alg:eliminate-variable}, each element of the user variable must be solved for independently. 
In this scenario, \equalitySolverSymbol{} is invoked on every sub-index of the tensor, resulting in a runtime of at least $(\prod_{i} s_i)(\sum_i \prod_{j = 0}^i s_i)$. 
If either the tensor has many dimensions or any dimension is large, this worst-case complexity quickly becomes infeasible~\footnote{We ignore here the additional exponential blow-up that can arise from disjunctions in the specification. This is reasonable, as in such cases compilation time is dominated by the cost of solving an exponential number of generated VNN-LIB queries.}.

It is important to emphasise that this behaviour reflects a pathological worst case. 
In practice, specifications arising in the neural network verification literature typically involve relatively few input dimensions and sparse equalities.
When dimensions are large (e.g. image inputs), specifications are usually written so that user variables can be eliminated at a higher level of the tensor structure.
In these common cases, the optimisations described in Algorithms~\ref{alg:elim-exists}~\&~\ref{alg:equality-elim} apply, and \equalitySolverSymbol{} is is invoked only a small number of times during Algorithm~\ref{alg:compile-body}. 
As a result, the effective runtime is close to $O(\prod_{i} s_i)$, i.e. linear in the total number of elements of the input tensor.

\paragraph{Runtime experiments} - To provide evidence for this claim, we implement the algorithm in the Vehicle framework~\cite{daggitt2025vehicle}\footnote{This Vehicle implementation supports a richer specification language than has been presented in this paper. A description of these extra features have been omitted due to space limitations. See Appendix~\ref{app:extensions} for details.}. We evaluate its performance on three common neural network specifications (available in the supplementary material): 
\begin{enumerate}
\item \textbf{ACASXu} - All~10 properties in the ACAS Xu specification~\cite{katz2017reluplex} - an example of a large, highly interpretable specification. Input tensor size is fixed.
\item \textbf{Robustness} - An encoding of standard $\epsilon$-ball robustness - an example of a specification where input tensor size can grow arbitrarily large.
\item \textbf{Monotonicity} - An encoding of monotonicity - an example of a specification with multiple network applications. Input tensor size can grow arbitrarily.
\end{enumerate}
The experiments were run on a 11th Gen Intel Core i7-1165G7 processor with 32 GiB RAM. The compilation time was measured by subtracting the time taken to type-check the specification from the mean compilation time for the specification over 10 runs.
For the robustness and monotonicity specifications, the number of elements in the input tensor tensors were varied between 100 and 6000.
\begin{figure*}[tpb]
	\begin{tikzpicture}
\begin{axis}[
    xlabel={Number of inputs to the network},
    ylabel style={align=center},
    ylabel={Mean compilation time\\over 10 runs (s)},
    xmin=0, xmax=7000,
    ymin=0, ymax=40,
    xtick={0,2000,4000,6000},
    ytick={0,10,20,30,40},
    width=\textwidth,
    height=\textwidth/2.3,
    legend style={at={(0.35,0.95)},anchor=north},
    ymajorgrids=true,
    grid style=dashed,
]
    
\addplot[
    color=blue,
    mark=square,
    error bars/.cd,
            y dir=both,y explicit,
    ]
coordinates {
    (100,0.2546) +- (0.0031,0.0031)
    (1000,1.866) +- (0.04,0.04)
    (2000,4.654) +- (0.13,0.13)
    (3000,8.931) +- (1.177,1.177)
    (4000,14.674) +- (2.01,2.01)
    (5000,18.43) +- (2.478,2.478)
    (6000,26.574) +- (3.478,3.478)
};
\addlegendentry{robustness}

\addplot[
    color=blue,
    mark=diamond,
    error bars/.cd,
            y dir=both,y explicit,
    ]
coordinates {
    (100,0.4626) +- (0.229,0.0128)
    (1000,3.791) +- (0.229,0.229)
    (2000,9.250) +- (1.052,01.052)
    (3000,14.599) +- (2.109,2.109)
    (4000,20.359) +- (1.575,1.575)
    (5000,23.363) +- (2.341,2.341)
    (6000,31.337) +- (2.602,2.602)
};
\addlegendentry{monotonicity}

\addplot[
    color=red,
    mark=square,
    error bars/.cd,
            y dir=both,y explicit,
    ]
coordinates {
    (100,0.747) +- (0.016,0.016)
    (200,2.277) +- (0.052,0.052)
    (300,5.156) +- (0.782,0.782)
    (400,10.820) +- (2.197,2.197)
    (500,16.085) +- (2.095,2.095)
    (600,20.712) +- (1.644,1.644)
    (700,30.604) +- (2.5,2.5)
    (800,47.790) +- (6.816,6.816)
};
\addlegendentry{robustness (ablation)}
\end{axis}
\end{tikzpicture}
    \vspace{-2em}
	\caption{Algorithm performance as the number of inputs to the network increases.}
	\label{fig:scaling-experiments}
\end{figure*}

The results are as follows. The entire ACAS Xu specification is compiled into 42 separate queries in $0.470 \pm 0.013$s.
Figure~\ref{fig:scaling-experiments} shows that the compilation time for the robustness and monotonicity specifications grow linearly.
This is asymptotically optimal, as the number of assertions in the resulting queries is also necessarily linear in the number of inputs to the network.
For those interested in further performance results, timings for a wide range of additional specifications can be found by running the Vehicle test suite.

\paragraph{Ablation experiments}  
To justify the careful optimisations presented, we performed a simple ablation study by adjusting Algorithm~\ref{alg:compile-comparison} to always reduce tensor comparisons to comparisons over elements. In this case Algorithms~\ref{alg:eliminate-variable}~\&~\ref{alg:equality-elim} are forced to always solve for equalities over tensor elements.
As can be seen in Figure~\ref{fig:scaling-experiments}, compilation time quickly becomes impractical for robustness specifications over even relatively small networks. This justifies our original assertions in Section~\ref{sec:motivating-example} about the difficulty of compiling to VNN-LIB vs SMT-LIB. 
\section{Conclusions}
\label{sec:conclusion}

We have described a numerically-sound algorithm for compiling specifications expressed in a high-level language into low-level VNN-LIB 2.0 queries consumable by neural network solvers. 
As discussed in Section~\ref{sec:existing-languages}, our source language is substantially more expressive than comparable high-level specification languages used in existing tools, while the compilation algorithm achieves asymptotically optimal performance for a broad class of commonly occurring specifications.

This increase in expressiveness is not merely theoretical; it enables practitioners to naturally encode specifications arising in realistic case studies that are awkward or impossible to express in existing frameworks.
For example, we encourage readers to consult the supplementary material and compare the high-level ACASXu specification with the compiled VNN-LIB queries. 
The ability to transform user variables before using them as network inputs means that the normalisation procedure can be defined as part of the specification, and the constraints can therefore be written in real world units (radians, m/s etc.).

This work therefore helps bridge the gap between human-readable intent and solver-level encodings and constitutes a key step toward the vision articulated in the recent community paper~\cite{cordeiro2025neural} of making neural network verification accessible beyond the research community.
By enabling non-expert users to write precise, solver-independent specifications, our approach lowers a major barrier to adoption and helps move neural network verification from a research field toward a scalable, trustworthy engineering discipline.

\bibliographystyle{splncs04}
\bibliography{bibliography.bib}

\appendix
\clearpage

\section{Additional algorithmic details}

\subsection{Implementing \orProtoqueriesSymbol{} and \andProtoqueriesSymbol{} over proto-queries}
\label{app:proto-query-logic}

As proto-queries are lists of implicitly disjuncted satisfaction problems, \orProtoqueriesSymbol{} can be implemented by concatenating lists of proto-queries, while \andProtoqueriesSymbol{} can be implemented by taking their Cartesian product and logically conjuncting each resulting pair.
After both operations, the resulting list is post-processed to merge proto-queries $p_1$ and $p_2$ that yield identical solutions for the user variables as functions of the network variables, replacing them with a single proto-query $\hOr{p_1}{p_2}$.
In order to implement this, it is necessary to augment the proto-queries data structure in Figure~\ref{fig:intermediate-syntax} with the solutions for the user variables in a suitable normal form. This is relatively trivial to do, and but a formal description of this data-structure and its calculation in Algorithm~\ref{alg:eliminate-variable} has been omitted due to space constraints.

\subsection{Definition of \evalAtSymbol{}}
\label{app:eval-lookup}

\begin{algorithm}[h]
    \begin{algorithmic}[1]
\newcommand{\dims}{\ensuremath{ds}}

\Require (i)~$e$ is a value of type $\hRealTensorType{\cons{d}{ds}}$ and $i$ is an index such that $i < d$, (ii)~$e$ is the output of \purifySymbol{} and therefore contains no network applications, variables or lookup operations at the current dimension.
\Ensure Outputs an arithmetic expression equivalent to the value of $e$ at index $i$.
\Function{\evalAtSymbol{}}{$e$, $i$}
    \Switch{$e$}
        \Case{$t$}
            \Return $t_i$
        \EndCase
        \Case{$\hAdd{e_1}{e_2}$}
            \Return \evalAt{$e_1$}{$i$} + \evalAt{$e_2$}{$i$}
        \EndCase
        \Case{$\hMul{e_1}{e_2}$}
            \Return \evalAt{$e_1$}{$i$} * \evalAt{$e_2$}{$i$}
        \EndCase
        \Case{$\hSeq{e_1 \ldots e_d}$}
            \Return $e_i$
        \EndCase
        \Case{$f e$}
            \Throw AssumptionViolated
        \EndCase
        \Case{$\hAt{e}{i}$}
            \Throw AssumptionViolated
        \EndCase
        \Case{$x$}
            \Throw AssumptionViolated
        \EndCase
    \EndSwitch
\EndFunction
\end{algorithmic}
    \caption{Eliminating an indexing operation}
    \label{alg:eval-at}
\end{algorithm}

Algorithm~\ref{alg:eval-at} shows the definition of the function \evalAtSymbol{} which was used in Algorithm~\ref{alg:compile-linear-expr} in Section~\ref{sec:properties}.
The key assumption is that the input tensor expression has already been passed through \purifySymbol{} and so all of the constructors that could block the evaluation of the indexing operation have already been eliminated.

\section{Extensions}
\label{app:extensions}

As mentioned in Section~\ref{sec:evaluation}, the Vehicle language supports many additional language features not present in the core calculus in this paper and the example specifications in the supplementary material make use of some of these features. 
Many of these features (e.g. quantifying over indices) are surface-level only and can be compiled into the core calculus given in Figure~\ref{fig:vehicle-syntax}. However, there are a few that cannot be compiled, and the core calculus and the algorithm itself must be extended.
None of them fundamentally change the essence of the algorithm, but some features require more subtle alterations than others. 
Below we provide sketches of how the Vehicle implementation extends the algorithm in various ways:
\begin{enumerate}
    \item \textbf{Boolean tensors}: The Vehicle language contains operations over boolean tensors, including reduction operations (e.g. \texttt{reduceAnd}, \texttt{reduceOr}), which also need to be eliminated by the compilation algorithm. Both \elimNotSymbol{} and \compileBodySymbol{} may encounter reduction operations which are blocked from evaluating. In this case it is necessary to perform a traversal of the blocking argument to find the tensor variables that are blocking reduction and replacing them with a stack of variables representing their sub-tensors, thereby allowing evaluation to continue. This new traversal function can therefore be seen as performing the same operation over Boolean expressions as \purifySymbol{} does for numeric expressions.

    \item \textbf{If statements}: The Vehicle language also contains \texttt{if-then-else} constructs. When the result of an \texttt{if} is a Boolean expression, then the \texttt{if-then-else} can be eliminated immediately by converting it into a disjunction when encountered by \elimNotSymbol{} and \compileBodySymbol{}. However when, as is common, the result of an \texttt{if} is a numeric expression then compilation is more complicated as the condition of the \texttt{if} may contain further quantifiers (see below). To handle this, it is first necessary to extend \purifySymbol{} to lift any \texttt{if} statements to the comparison level. For example, 
    \begin{center}
    \texttt{0 < (if (exists x . f x > 0) then f y else f [y + 1])}        
    \end{center}
    becomes
    \begin{center}
    \texttt{if (exists x . f x > 0) then 0 < f y else 0 < f [y + 1]}        
    \end{center}
    When such an \texttt{if} is detected and lifted, then \compileComparisonSymbol{} must recursively invoke \compileBodySymbol{}.
    
    \item \textbf{Networks with multiple inputs and outputs}: Multi-modal neural networks are becoming increasingly popular, but the algorithm presented in this paper assumes the network has a single input and output tensor. To handle this, the following modifications can be made:
    \begin{enumerate}
        \item the high-level language should be extended with records, with each field representing one of the input/output tensors.
        \item the hierarchy of variables in Figure~\ref{fig:intermediate-syntax} must be extended to also assign a variable to each record field.
        \item The neural network application case in \ref{alg:compile-linear-expr} must be altered to introduce multiple input equalities rather than a single input equality.
        \item The \findConnectedComponentsSymbol{} algorithm must be altered to ensure that two network applications are judged equal only if all their inputs are equal. 
    \end{enumerate}
\end{enumerate}

\end{document}